\documentclass[12p]{article}

\begin{document}

\input amssym.tex

\title{New Dirac quantum modes in moving frames of the
de Sitter spacetime}

\author{Ion I.  Cot\u{a}escu\thanks{E-mail:~~~cota@physics.uvt.ro}\,
        and Cosmin Crucean\thanks{E-mail:~~~crucean@physics.uvt.ro }\\
{\small \it West University of Timi\c soara,}\\
       {\small \it V.  P\^ arvan Ave.  4, RO-300223 Timi\c soara, Romania}}

\maketitle

\begin{abstract}
Recently a new time-evolution picture of the Dirac quantum mechanics  was
defined in charts with spatially flat Robertson-Walker metrics, under the name
of Schr\"{o}dinger picture [I. I. Cot\u{a}escu,  gr-qc/0708.0734] . In the
present paper new Dirac quantum modes are found in moving charts of the de
Sitter spacetime using the technical advantages offered by this picture. The
principal result is a new set of energy eigenspinors  which behave as polarized
plane waves and form a complete system of orthonormalized solutions of the free
Dirac equation.

 Pacs:
04.62.+v
\end{abstract}

\newpage

\section{Introduction}

The quantum modes of the free Dirac field on de Sitter (dS) backgrounds is
well-studied in different local charts (or natural frames) and tetrad gauge
fixings.

The first solutions of the free Dirac equation on dS backgrounds are the energy
eigenspinors found by Otchik \cite{OT} using the diagonal gauge in central
charts with spherical coordinates (i.e. static charts with spherical symmetry).
A special type of rotation-covariant Cartesian gauge in central charts allowed
us to express these eigenspinors in terms of the well-known  spherical spinors
of special relativity \cite{co}. Other interesting solutions in these charts
were derived recently under the null tetrad gauge \cite{LO}.

In  dS spacetimes there are  moving charts equipped with Cartesian or spherical
coordinates whose line elements are of the Robertson-Walker (RW) type. The
first spherical wave solutions of the free Dirac equation in moving charts with
spherical coordinates were obtained by Shishkin  \cite{SHI}. We have shown that
some linear combinations of these solutions are eigenspinors of the scalar
momentum operator  \cite{co1}. Other solutions are the polarized plane waves
derived as eigenspinors of the momentum operator in moving frames with
Cartesian coordinates and diagonal gauge  \cite{BFGF}. We have correctly
normalized these solutions in the momentum scale  pointing out that these
constitute a complete system of orthonormalized spinors \cite{cot}. With their
help we quantized the Dirac field in  canonical manner writing down the
expressions of the most important conserved operators of the field theory
\cite{cot}.

Thus we see that in moving frames we know only Dirac modes with well-defined
momentum but in which the energy can not be measured exactly since the momentum
operators do not commute with the Hamiltonian one \cite{cot}. Consequently, an
important non-trivial problem remains open, namely that of finding the Dirac
{\em energy} eigenspinors in moving charts of the dS spacetimes. In this paper
we would like to solve this problem using the theory of time evolution pictures
of the Dirac quantum mechanics in spatially flat RW geometries we have recently
proposed \cite{cor}.

In the non-relativistic quantum mechanics the time evolution can be studied in
different pictures (e.g. Schr\" odinger, Heisenberg, etc.) which transform
among themselves through  specific time-dependent unitary transformations. In
special and general relativity, despite of its importance, the problem of
time-evolution pictures is less studied because of some technical difficulties
related to the Klein-Gordon equation which has no Hamiltonian form. However,
the Dirac quantum mechanics is a convenient framework for introducing different
pictures as auxiliary tools since the Dirac equation can be brought in
Hamiltonian form at any time. We have shown that at least two time evolution
pictures can be identified in the case of the Dirac theory on backgrounds with
spatially flat RW metrics. We considered that the {\em natural} picture (NP) is
that in which the free Dirac equation is written directly as it results from
its Lagrangean, in a diagonal gauge and Cartesian coordinates \cite{cot}. The
second one, called the Schr\" odinger picture (SP), is a new picture where the
free Dirac equation is transformed such that its kinetic part takes  the same
form as in special relativity while the gravitational interaction is separated
in a specific term. In this picture we defined the principal operators of our
theory obtaining thus the ingredients we need for determining quantum modes
\cite{cor}.

Here we use the SP for deriving new Dirac quantum modes in moving frames of the
dS spacetime. First we study new energy eigenspinors which behave as spherical
waves in moving frames but describing similar quantum modes as those derived
previously in cental frames \cite{co}. Furthermore, we focus on quite new Dirac
quantum modes in moving frames whose spinors  are polarized plane waves
solutions of the free Dirac equation determined by energy, momentum direction
and helicity. These plane waves can be normalized in the energy scale (in
generalized sense) and form a {\em complete} system of energy eigenspinors.

We start in the second section with a brief review of the Dirac quantum
mechanics in spatially flat RW backgrounds, including our theory of time
evolution pictures. In the next section we derive the mentioned new Dirac
quantum modes in moving frames of the dS spacetime  where the Dirac equation in
SP is analytically solvable either in coordinates or even in momentum
representation. The spherical waves are obtained separating the spherical
coordinates while the polarized plane waves are derived separating variables in
momentum representation. The orthonormalization and completeness properties of
the plane wave solutions are also deduced.

\section{Dirac fields in spatially flat RW spacetimes}

In what follows we present the principal time evolution pictures of the Dirac
quantum mechanics on spatially flat RW backgrounds, pointing out the technical
advantages of using different pictures.

\subsection{The Dirac equation in NP}

The relativistic quantum mechanics we discuss here is build in  local chart
with Cartesian or spherical coordinates, $x^{\mu}$ ($\mu,\nu,...=0,1,2,3 $), of
a $(1+3)$-dimensional spatially flat RW manifold. These are the proper time
$x^0=t$ and either the Cartesian space coordinates $x^i$ ($i,j,...=1,2,3$) or
the associated spherical ones, $r=|\vec{x}|$, $\theta$ and $\phi$. In these
charts the RW line element
\begin{equation}\label{line}
ds^2=dt^2-\alpha(t)^2 (d\vec{x}\cdot
d\vec{x})=dt^2-\alpha(t)^2(dr^2+r^2d\theta^2+ r^2\sin^2\theta d\phi^2)
\end{equation}
depends only on the arbitrary function $\alpha$.

In gauge-covariant field theories, the fields with spin half obey equations
whose form strongly depends on the choice of the local orthogonal frames and
coframes with respect of which the spin operators are defined. The local frames
and coframes are given by the tetrad fields $e_{\hat\mu}(x)$ and, respectively,
$\hat e^{\hat\mu}(x)$ \cite{SW}. These fields are labeled by the local indices
($\hat\mu,\hat\nu,...=0,1,2,3$) of the Minkowski metric
$\eta=$diag$(1,-1,-1,-1)$, satisfy $e_{\hat\mu}^{\alpha}\hat
e^{\hat\mu}_{\beta}=\delta^{\alpha}_{\beta}$, $e_{\hat\mu}^{\alpha}\hat
e^{\hat\nu}_{\alpha}=\delta^{\hat\mu}_{\hat\nu}$  and give the metric tensor as
$g_{\mu \nu}=\eta_{\hat\alpha\hat\beta}\hat e^{\hat\alpha}_{\mu}\hat
e^{\hat\beta}_{\nu}$.

The free Dirac field $\psi$ of mass $m$, seen as a perturbation that does not
affect the geometry, satisfies the free Dirac equation which can be easily
written in the diagonal gauge where the tetrad fields have the non-vanishing
components \cite{BFGF,SHI},
\begin{equation}\label{tt}
e^{0}_{0}=1\,, \quad e^{i}_{j}=\frac{1}{\alpha(t)}\delta^{i}_{j}\,,\quad \hat
e^{0}_{0}=1\,, \quad \hat e^{i}_{j}=\alpha(t)\delta^{i}_{j}\,.
\end{equation}
In this gauge one obtains the usual form of the Dirac equation in RW spacetimes
\cite{BFGF},
\begin{equation}\label{ED1}
\left(i\gamma^0\partial_t+i\frac{1}{\alpha(t)}\gamma^i\partial_i
+\frac{3i}{2}\frac{\dot{\alpha}(t)}{\alpha(t)}\gamma^{0}-m\right)\psi(x)=0\,,
\end{equation}
expressed in terms of Dirac $\gamma$-matrices \cite{TH}, with the notation
$\dot{\alpha}=\partial_t\alpha$. The Dirac quantum modes are described by
solutions of Eq. (\ref{ED1}) that behave as tempered distributions or square
integrable spinors with respect to the relativistic scalar product \cite{BD},
\begin{equation}\label{SP1}
\left<\psi,\psi'\right>=\int_D d^3x
\sqrt{g(t)}\,\bar{\psi}(x)\gamma^0\psi'(x)\,,
\end{equation}
where $\sqrt{g(t)}=\sqrt{|\det g_{\mu\nu}(t)|}=\alpha(t)^3$ play the role of a
weight function. The field $\bar{\psi}=\psi^+\gamma^0$ is the Dirac adjoint of
$\psi$ and $D$ is the space domain of the chart we use.

The above choice of the local chart and tetrad fields leads to the NP in which
the time evolution is governed by the Dirac equation (\ref{ED1}) resulted from
the  standard formalism of the gauge-covariant field theories, without other
transformations or artifices. The principal operators of this picture, the
energy $\hat H$, momentum $\vec{\hat P}$ and coordinate $\vec{\hat X}$, can be
defined as in special relativity,
\begin{equation}\label{ON}
(\hat H \psi)(x)=i\partial_t\psi(x)\,,\quad (\hat P^i
\psi)(x)=-i\partial_i\psi(x)\,,\quad (\hat X^i \psi)(x)=x^i\psi(x)\,.
\end{equation}
The operators $\hat X^i$ and $\hat P^i$ are time-independent and satisfy the
well-known canonical commutation relations
\begin{equation}\label{com}
\left[\hat X^i, \hat P^j\right]=i\delta_{ij}I\,,\quad \left[\hat H, \hat
X^i\right]=\left[\hat H,\hat P^i\right]=0\,,
\end{equation}
where $I$ is the identity operator. Other operators are formed by orbital parts
and suitable spin parts that may be point-dependent too. In general, the
orbital terms are freely generated by the basic orbital operators $\hat X^i$
and $\hat P^i$. An example is the total angular momentum
$\vec{J}=\vec{L}+\vec{S}$ where $\vec{L}=\vec{\hat X}\times\vec{\hat P}$ and
$\vec{S}$ is the spin operator. We specify that the operators $\hat P^i$ and
$J^i$ are generators of the spinor representation of the group $\tilde E(3)$
defined as the universal covering group of the isometry group $E(3)$ of the
spatially flat RW manifolds \cite{cot}. Therefore, these operators are
 conserved in the sense that they commute with the Dirac operator
\cite{CML,ES}.

\subsection{The Dirac equation in SP}

The NP can be changed using point-dependent operators which could be even
non-unitary operators since the relativistic scalar product does not have a
direct physical meaning as that of the non-relativistic quantum mechanics. We
exploited this opportunity for defining the SP in coordinate representation
\cite{cor} but in this picture the momentum representation is also efficient
for studying quantum modes.

Let us start with the coordinate representation where we defined the SP as the
picture in which the kinetic part of the Dirac operator takes the standard form
$i\gamma^0\partial_t+i\gamma^i\partial_i$ \cite{cor}. The transformation
$\psi(x)\to \psi_S(x)=W(x)\psi(x)$ leading to the SP is produced by the
operator of time dependent dilatations
\begin{equation}\label{U}
W(x)=\exp\left[-\ln(\alpha(t))(\vec{x}\cdot\vec{\partial})\right]\,,
\end{equation}
which has the remarkable property \footnote {We denote by  $(~)^{\dagger}$ the
adjoint operators with respect to the scalar product (\ref{SP1}) and by
$(~)^{+}$ the adjoint matrices.}
\begin{equation}\label{Udag}
W(x)^{\dagger}= \sqrt{g(t)}\, W(x)^{-1} \,,
\end{equation}
and the following convenient action
\begin{equation}
W(x)F(\vec{x})W(x)^{-1}=F\left(\frac{1}{\alpha(t)}\vec{x}\right)\,,\quad
W(x)G(\vec{\partial})W(x)^{-1}=G\left(\alpha(t)\vec{\partial}\right)\,,
\end{equation}
upon any analytical functions $F$ and $G$. Performing this transformation we
obtain the free Dirac equation of SP
\begin{equation}\label{ED2}
\left[i\gamma^0\partial_{t}+i\vec{\gamma}\cdot\vec{\partial} -m
+i\gamma^{0}\frac{\dot{\alpha}(t)}{\alpha(t)}
\left(\vec{x}\cdot\vec{\partial}+\frac{3}{2}\right)\right]\psi_S(x)=0\,,
\end{equation}
and the new form of the relativistic scalar product,
\begin{equation}\label{SP2}
\left<\psi_S,\psi'_S\right>=\left<\psi,\psi'\right>= \int_D d^3x
\,\bar{\psi}_S(x)\gamma^0\psi_S'(x)\,,
\end{equation}
calculated from Eqs. (\ref{SP1}) and (\ref{Udag}). We observe that this is no
more dependent on $\sqrt{g(t)}$, taking the same form from as in special
relativity.

The specific operators of SP, denoted by $H_S$, $P^i_S$ and $X^i_S$, are
defined in usual manner as
\begin{equation}\label{OS}
(H_S \psi_S)(x)=i\partial_t\psi_S(x)\,,~~ (P^i_S
\psi_S)(x)=-i\partial_i\psi_S(x)\,,~~ (X^i_S \psi_S)(x)=x^i\psi_S(x)\,,
\end{equation}
obeying commutation relations similar to Eqs. (\ref{com}). With their help the
Dirac equation (\ref{ED2}) can be put in Hamiltonian form,
\begin{equation}\label{Ham}
i\partial_t \psi_S(x)={\cal H}_S\psi_S(x)\,
\end{equation}
where the Dirac Hamiltonian operator ${\cal H}_S={\cal H}_0 + {\cal H}_{int}$
has the standard kinetic term ${\cal H}_0=\gamma^0\vec{\gamma}\cdot
\vec{P}_S+\gamma^0 m$ and the interaction term with the gravitational field,
\begin{equation}\label{Hint}
{\cal H}_{int}=\frac{\dot{\alpha}(t)}{\alpha(t)}\left(\vec{X}_S\cdot
\vec{P}_S-\frac{3i}{2}I\right)=\frac{\dot{\alpha}(t)}{\alpha(t)}\left(\vec{\hat
X}\cdot \vec{\hat P}-\frac{3i}{2}I\right)\,,
\end{equation}
that is proportional just to the Huble function $\dot{\alpha}/\alpha$. In these
circumstances, we assume that the correct quantum observables are the operators
defined by Eqs. (\ref{OS}). Performing the inverse transformation we find that
in NP these operators become new interesting time-dependent operators,
\begin{eqnarray}
H(t)&=&W(x)^{-1}H_SW(x)=\hat H+\frac{\dot{\alpha}(t)}{\alpha(t)} \vec{\hat
X}\cdot\vec{\hat P}\,,\label{Ht} \\
X^i(t)&=&W(x)^{-1}X_S^iW(x)=\alpha(t) \hat X^i\,,\\
P^i(t)&=&W(x)^{-1}P_S^iW(x)=\frac{1}{\alpha(t)}\hat P^i\,,\label{Pt}
\end{eqnarray}
which satisfy usual commutation relations as those given by Eqs. (\ref{com}).
The angular operators, $\vec{J}$ and $K$, as well as the operator (\ref{Hint})
have the same expressions in both these pictures since they commute with
$W(x)$.

In NP the eigenvalues problem $H(t)f_E(t,\vec{x})=Ef_E(t,\vec{x})$ of the
Hamiltonian operator (\ref{Ht}) leads to energy eigenfunctions of the form
\begin{equation}
f_E(t,\vec{x})=F[\alpha(t)\vec{x}]e^{-iEt}
\end{equation}
where $F$ is an arbitrary function. This explains why in this picture one can
not find energy eigenstates separating  variables. However, in SP these
eigenfunctions become the new functions
\begin{equation}
f^S_E(t,\vec{x})=W(x)f_E(t,\vec{x})=F(\vec{x})e^{-iEt}
\end{equation}
which have separated variables. This means that in  SP new quantum modes could
be derived using the method of separating variables in coordinates or even in
momentum representation.

\section{New energy eigenspinors in moving frames of dS spacetimes}

Now we shall see how can be used the SP for solving the mentioned problem of
finding Dirac energy eigenspinors in  moving frames of the dS spacetime. There
are two types of solutions of the Dirac equation, determined by different sets
of commuting operators which include the Hamiltonian operator. We show that
these behave either as spherical waves or as polarized plane waves.

\subsection{New spherical waves}

In the particular case of the dS spacetime there is a moving chart
$\{t,r,\theta,\phi\}$ with spherical coordinates which has the line element
(\ref{line}) with $\alpha(t)=e^{\omega t}$. Another important chart with
spherical coordinates is the central chart $\{t_s,r_s,\theta,\phi\}$ having the
line element
\begin{equation}
ds^2=(1-\omega^2 r_s^2)dt_s^2-\frac{dr^2_s}{1-\omega^2
r_s^2}-r^2_s(d\theta^2+\sin^2\theta\, d\phi^2)\,.
\end{equation}
In each of these charts the observers which stay at $\vec{x}=0$ have specific
event horizons. Thus an observer $A$ situated at the point $r=0$ of the moving
chart and following the trajectory of $\partial_t$ has an event horizon at
$r=1/\omega$. In the central chart, another observer, $B$, at $r_s=0$ with the
trajectory along to the direction of the Killing vector field $\partial_{t_s}$,
can observe events up to $r_s=1/\omega$ \cite{BD}. This means that the radial
domain of the moving chart where we have to investigate quantum properties may
be $D_r=[0,1/\omega)$.

These observers measure quantum modes of the free Dirac field using not only
different coordinates but different local frames too. We assume that $B$
observes the quantum modes of Ref. \cite{co}, using the Cartesian gauge defined
therein, while in the moving frame the observer $A$ chooses the present
diagonal gauge (\ref{tt}). These tetrad gauge fixings are different since the
time axis of the local frame of $B$ is along to $\partial_{t_s}$ but the time
axis of the local frame of $A$ is along to  $\partial_t$.  However, despite of
these differences, both these observers recognize the same physical quantities
globally defined as conserved operators produced by the Killing vectors of the
dS geometry. The Hamiltonian operator (\ref{Ht}) is just the conserved operator
$H=i\partial_{t_s}$ produced by the time-like Killing vector. In dS moving
frames and NP this has the form $H=i\partial_t+\omega \vec{\hat
X}\cdot\vec{\hat P}$ \cite{cot} while in SP its action is given by the first of
Eqs. (\ref{OS}). Other conserved operators are the components of the total
angular momentum $\vec{J}$ which has the same form for both the observers if
they keep unchanged the gauge fixings specified above \cite{ES,cot}. We note
that the operators $\hat P^i$ are also conserved and more three other operators
fill out the set of ten conserved generators of the spinor representation of
the universal covering group of the dS isometry group $SO(1,4)$ \cite{cot,ES}.

Our purpose is to derive the Dirac energy eigenspinors  observed by $A$ as
spherical waves in dS moving frames.  This can be achieved only in SP where the
free Dirac equation,
\begin{equation}\label{ED3}
\left[i\gamma^0\partial_{t}+i\vec{\gamma}\cdot\vec{\partial} -m
+i\gamma^{0}\omega
\left(\vec{x}\cdot\vec{\partial}+\frac{3}{2}\right)\right]\psi_S(x)=0\,,
\end{equation}
does not depend explicitly on time. Therefore, in this picture we can separate
the spherical variables of the Dirac equation as in the problems with spherical
symmetry of special relativity. The starting point is the Dirac equation put in
the form (\ref{Ham}) with the Dirac Hamiltonian operator of the SP written in
terms of radial and angular operators as,
\begin{equation}\label{HHH}
{\cal H}_S=-\frac{i}{r^2}\left[\gamma^0(\vec{\gamma}\cdot\vec{x})\left(
\vec{x}\cdot\vec{\partial}+1\right)+(\vec{\gamma}\cdot\vec{x})\,K\right]
+\gamma^0 m-i\omega \left(\vec{x}\cdot\vec{\partial}+\frac{3}{2}\right)\,,
\end{equation}
where $K=\gamma^0(2\vec{L}\cdot\vec{S}+1)$ is the Dirac angular operator.
Furthermore, we have to look for particular solutions of the Dirac equation
defined as common eigenspinors of the complete set of commuting operators
$\{{\cal H}_S, \vec{J}^2, K, J_3\}$ corresponding to the set of eigenvalues
$\{E,j(j+1),-\kappa_j,m_j\}$ formed by energy, $E$, angular quantum numbers,
$j$ and $m_j$, and $\kappa_j=\pm(j+1/2)$ \cite{TH}. The particular solutions of
particle type are the positive frequency spinors,
\begin{eqnarray}
&&U^S_{E,\kappa_j,m_j}(t,r,\theta,\phi)\nonumber\\
&&~~~~~~=\frac{1}{r}\,\left[
f^{(+)}_{E,\kappa_j}(r)\Phi^{+}_{m_{j},\kappa_{j}}(\theta,\phi) +
f^{(-)}_{E,\kappa_j}(r)\Phi^{-}_{m_{j},\kappa_{j}}(\theta,\phi)\right]e^{-iEt}\,,\label{psol}
\end{eqnarray}
which depend on a pair of radial functions, $f^{(\pm)}_{E,\kappa_j}$, and the
usual spherical spinors $\Phi^{\pm}_{m_j,\kappa_j}$ which completely solve the
angular eigenvalues problem \cite{TH}. Taking into account that these spinors
are orthonormalized with respect to the angular scalar product \cite{TH} we
find that the scalar product of two spinors of the form (\ref{psol}) reads
\begin{eqnarray}
&&\left<U^S_{E,m_j,\kappa_j},U^S_{E',m'_j,\kappa'_j}\right>
=\delta_{\kappa_j,\kappa'_j}\delta_{m_j,m'_j}\nonumber \\
&&~~~~~~~~~~\times
\int_{D_r}dr\left\{[f^{(+)}_{E,\kappa_j}(r)]^*f^{(+)}_{E',\kappa_j}(r)+
[f^{(-)}_{E,\kappa_j}(r)]^*f^{(-)}_{E',\kappa_j}(r)\right\}\,.\label{pp}
\end{eqnarray}

After the separation of the angular variables of the Dirac equation we are left
with the pair of radial equations
\begin{equation}\label{Sys}
\left(\pm m-E-\frac{i\omega}{2}-i \omega r \frac{d}{dr}\right)
f^{(\pm)}_{E,\kappa_j}(r) =\left(\pm\frac{d}{dr}-\frac{\kappa_j}{r}\right)
f^{(\mp)}_{E,\kappa_j}(r)
\end{equation}
that can be rewritten as
\begin{eqnarray}
\left[(1-\omega^2r^2)\frac{d}{dr}\pm \frac{\kappa_j}{r}+ i\omega r\left(E\mp
m+\frac{i\omega}{2}\right)\right]f^{(\pm)}_{E,\kappa_j}(r)~\,&&\nonumber\\
=\left[\pm E+m+i\omega\left(\kappa_j\pm
\frac{1}{2}\right)\right]f^{(\mp)}_{E,\kappa_j}(r)\,.&&
\end{eqnarray}
This system has to be analytically solved in the radial domain $D_r$ looking
for tempered distributions corresponding to a continuous energy spectrum. This
means that these solutions  should not have singularities at $r=0$. When
$\kappa_j=j+\frac{1}{2}>0$ we find regular solutions of the form
\begin{equation}
f^{(\pm)}_{E,\kappa_j}(r)=N^{(\pm)}_{E,\kappa_j}(\omega r)^{j+1\pm \frac{1}{2}}
F\left(a_{E,j}\,,\,b_{E,j}\pm \textstyle{\frac{1}{2}}\,;\,j+\frac{3}{2}\pm
\frac{1}{2}\,;\,\omega^2 r^2\right)\,,
\end{equation}
where the Gauss hypergeometric functions $F$ depend on the parameters
\begin{eqnarray}
a_{E,j}&=&-\frac{i}{2\omega}(E-m)+\frac{j}{2}+1\,,\\
b_{E,j}&=&-\frac{i}{2\omega}(E+m)+\frac{j}{2}+1\,,
\end{eqnarray}
while the normalization factors obey the condition
\begin{equation}
\frac{N^{(+)}_{E,\kappa_j}}{N^{(-)}_{E,\kappa_j}}=\frac{E+m}{2(j+1)}+\frac{i\omega}{2}\,.
\end{equation}
In the case of  $\kappa_j=-(j+\frac{1}{2})<0$  the regular solutions are
\begin{equation}
f^{(\pm)}_{E,-|\kappa_j|}(r)=N^{(\pm)}_{E,-|\kappa_j|}(\omega r)^{j+1\mp
\frac{1}{2}} F\left(b_{E,j}\,,\,a_{E,j}\mp
\textstyle{\frac{1}{2}}\,;\,j+\frac{3}{2}\mp \frac{1}{2}\,;\,\omega^2
r^2\right)\,,
\end{equation}
having normalization factors which satisfy
\begin{equation}
\frac{N^{(-)}_{E,-|\kappa_j|}}{N^{(+)}_{E,-|\kappa_j|}}=-\frac{E-m}{2(j+1)}+\frac{i\omega}{2}\,.
\end{equation}
All these solutions are regular on the domain $D_r$ but for $r\to 1/\omega$
these are divergent. Indeed, the parameters of the above hypergeometric
functions, $F(a,b;c;\omega^2r^2)$, do not satisfy the condition $\Re (c-a-b)>0$
which assures the convergence of these functions for $\omega r\to 1$ \cite{AS}.
The next step might be the normalization in the energy scale with respect to
the scalar product (\ref{pp}) but, unfortunately, this can not be done since
the radial functions are too complicated.

The results obtained in SP must be rewritten in NP where our observer $A$
measures the energy eigenstates. This will see the particle-like energy
eigenspinors
\begin{equation}
U_{E,\kappa_{j},m_j}(t,r,\theta,\phi)=W(x)^{-1}U^S_{E,\kappa_{j},m_j}(t,r,\theta,\phi)=
U^S_{E,\kappa_{j},m_j}(t,e^{\omega t}r,\theta,\phi)\,,
\end{equation}
which have no longer separated variables. The antiparticle energy eigenspinors
can be derived directly using the charge conjugation as in \cite{cot}. All
these spinors are regular in the domain where the condition $r_s=re^{\omega
t}<1/\omega$ is fulfilled.

It is important to specify that the quantum modes we derived above
(observed by $A$) and those reported in Ref. \cite{co}  (measured by
$B$) are {\em equivalent}, since in both these cases the solutions
of the Dirac equation are eigenspinors of the {\em same} system of
commuting operators, $\{{H}, \vec{J}^2, K, J_3\}$. For this reason
these modes are defined on the same domain of the dS manifold. In
this situation, one may ask how these observers could compare their
results, concluding that they measure the same quantum modes, when
they are using different coordinates and local frames. Obviously, a
transformation between these two types of modes must exist but this
can not be reduced to a simple coordinate transformation since one
has to change simultaneously the positions of the local frames. This
can be done with the help of a {\em combined} transformation
involving, beside the coordinate transformation, a suitably
associated gauge transformation \cite{ES}. Our preliminary
calculations indicate that this is a local Lorenz boost of parameter
$ {\rm argtanh}(\omega r_s)$, along the direction of $\vec{x}$.
However, details concerning this topics will be discussed elsewhere.

\subsection{New polarized plane waves}

The above arguments show that the spherical waves we obtained here describe in
fact quantum modes discovered formerly in central charts. Therefore, if this
would be the only new result due to our SP then the effort of studying new time
evolution pictures might appear as useless or at most of academic interest.
Fortunately, other quite new energy eigenspinors can be found in this
framework. We show that these are polarized plane wave solutions in helicity
basis that can be derived in momentum representation.

For solving the Dirac equation (\ref{ED3}) in momentum representation we assume
that the spinors of the SP may be expanded in terms of plane waves of positive
and negative frequencies as,
\begin{eqnarray}
&&\psi_S(x)=\psi^{(+)}_S(x)+\psi^{(-)}_S(x)\nonumber\\
&&=\int_0^{\infty}\,dE\int_{\hat D}\, d^3p\,\,
\left[\hat{\psi}^{(+)}_S(E,\vec{p})\,e^{-i(Et-\vec{p}\cdot \vec{x})}
+\hat{\psi}^{(-)}_S(E,\vec{p})\,e^{i(Et-\vec{p}\cdot \vec{x})}\right]\label{PsiS}
\end{eqnarray}
where $\hat{\psi}^{(\pm)}_S$ are spinors which behave as tempered distributions
on the domain $\hat{D}={\Bbb R}_p^3$ such that the Green theorem may be used.
Then we can replace the momentum operator $\vec{P}_S$ by  $\vec{p}$ and the
coordinate operator $\vec{X}_S$ by $i \vec{\partial}_p$ obtaining the free
Dirac equation of the SP in momentum representation,
\begin{equation}\label{ED4}
\left[\pm E\gamma^0\mp\vec{\gamma}\cdot\vec{p} -m -i\gamma^{0}\omega
\left(\vec{p}\cdot\vec{\partial}_p+\frac{3}{2}\right)\right]\hat{\psi}^{(\pm)}_S(E,\vec{p})=0\,,
\end{equation}
where $E$ is the energy defined as the eigenvalue of $H_S$. Denoting
$\vec{p}=p\, \vec{n}$ with $p=|\vec{p}\,|$, we observe that the differential
operator of Eq. (\ref{ED4}) is of radial type and reads
$\vec{p}\cdot\vec{\partial}_p=p\,\partial_p$. Therefore, this operator acts on
the functions which depend on $p$
 while the functions which depend only on the momentum direction $\vec{n}$
 behave as constants.

Following the method of Ref. \cite{cot}, we should like to derive the
fundamental solutions in {\em helicity} basis using the standard representation of the
$\gamma$-matrices (with diagonal $\gamma^0$) \cite{TH}. We start with
the general solutions,
\begin{eqnarray}
&&\hat{\psi}^{(+)}_S(E,\vec{p})=\sum_{\lambda}u^S({E,\vec{p},\lambda})\,a(E,\vec{n},\lambda)\,,\\
&&\hat{\psi}^{(-)}_S(E,\vec{p})=\sum_{\lambda}v^S({E,\vec{p},\lambda})\,b^*(E,\vec{n},\lambda)\,,
\end{eqnarray}
involving spinors of helicity $\lambda=\pm\frac{1}{2}$ and the wave functions $a$
and $b^*$ which play the role of constants since they do not
depend on $p$. According to our previous results \cite{cot}, the spinors of
the momentum representation must have the form
\begin{eqnarray}
u^S(E,\vec{p},\lambda)&=&\left(
\begin{array}{c}
\frac{1}{2}\,f^{(+)}_E(p)\,\xi_{\lambda}(\vec{n})\\
\lambda g^{(+)}_E(p)\,\xi_{\lambda}(\vec{n})
\end{array}\right)\,,\label{uS}\\
v^S(E,\vec{p},\lambda)&=&\left(
\begin{array}{c}
-\lambda g^{(-)}_E(p)\,\eta_{\lambda}(\vec{n})\\
\frac{1}{2}\,f^{(-)}_E(p)\,\eta_{\lambda}(\vec{n})
\end{array}\right)\,,\label{vS}
\end{eqnarray}
where $\xi_{\lambda}(\vec{n})$ and $\eta_{\lambda}(\vec{n}) =i\sigma_2
[\xi_{\lambda}(\vec{n})]^{*}$ are the Pauli spinors of helicity basis which
satisfy the eigenvalue equations
\begin{equation}
(\vec{n}\cdot \vec{\sigma})\,\xi_{\lambda}(\vec{n})
=2\lambda\xi_{\lambda}(\vec{n})\,,\quad (\vec{n}\cdot
\vec{\sigma})\,\eta_{\lambda}(\vec{n}) =-2\lambda\eta_{\lambda}(\vec{n})\,,
\end{equation}
and the othonormalization conditions
\begin{equation}\label{Porto}
[\xi_{\lambda}(\vec{n})]^+\xi_{\lambda'}(\vec{n})
=[\eta_{\lambda}(\vec{n})]^+\eta_{\lambda'}(\vec{n}) =\delta_{\lambda\,
\lambda'}\,.
\end{equation}
It remains to derive the radial functions solving the system
\begin{eqnarray}
&&\left[i\omega \left(p\,\frac{d}{dp}
+\frac{3}{2}\right)\mp(E-m)\right]f^{(\pm)}_E(p)=\mp p\, g^{(\pm)}_E(p)\,,\\
&&\left[i\omega \left(p\,\frac{d}{dp}
+\frac{3}{2}\right)\mp(E+m)\right]g^{(\pm)}_E(p)=\mp p\, f^{(\pm)}_E(p)\,,
\end{eqnarray}
resulted from Eq. (\ref{ED4}). Denoting  by $k=\frac{m}{\omega}$ and
$\epsilon=\frac{E}{\omega}$, we find the solutions
\begin{eqnarray}
&&f^{(+)}_E(p)=[-f^{(-)}_E(p)]^*=C p^{-1-i\epsilon} e^{\pi k/2}
H^{(1)}_{\nu_-}(\textstyle{\frac{p}{\omega}})\,,\label{fpus}\\
&&g^{(+)}_E(p)=[-g^{(-)}_E(p)]^*=C p^{-1-i\epsilon}e^{-\pi k/2}
H^{(1)}_{\nu_+}(\textstyle{\frac{p}{\omega}})\,,\label{gpus}
\end{eqnarray}
expressed in terms of Hankel functions of indices $\nu_{\pm}=\frac{1}{2}\pm
ik$ whose properties are briefly presented in the Appendix A.
The normalization constant $C$ has to assure the normalization in the energy scale.

Collecting all the above results we can write down the final
expression of the Dirac field (\ref{PsiS}) as
\begin{eqnarray}
\psi_S(x)=\int_0^{\infty}\,dE\int_{S^2}\, d\Omega_n\,\,
\sum_{\lambda}\left[U^S_{E,\vec{n},\lambda}(t,\vec{x}) a(E,\vec{n},\lambda)
\right.~~~&&\nonumber\\
\left.+\, V^S_{E,\vec{n},\lambda}(t,\vec{x}) b^*(E,\vec{n},\lambda)\right]\,,&&
\end{eqnarray}
where the integration covers the sphere $S^2\subset \hat D$. The notation $U^S$
and $V^S$ stands for the {\em fundamental} spinor solutions of positive and,
respectively, negative frequencies, with energy $E$, momentum direction
$\vec{n}$ and helicity $\lambda$. According to Eqs. (\ref{uS}), (\ref{vS}),
(\ref{fpus}) and (\ref{gpus}), these are
\begin{eqnarray}
&&U^S_{E,\vec{n},\lambda}(t,\vec{x})
\nonumber\\
&&~~~~= i Ne^{-iEt}\int_{0}^{\infty} s\, ds \left(
\begin{array}{c}
\frac{1}{2}\,e^{\pi k/2}H^{(1)}_{\nu_{-}}(s) \,
\xi_{\lambda}(\vec{n})\\
\lambda\, e^{-\pi k/2}H^{(1)}_{\nu_{+}}(s) \,\xi_{\lambda}(\vec{n})
\end{array}\right)
e^{i \omega s \vec{n}\cdot\vec{x}-i\epsilon \ln s}\,,\label{Ups}\\
&&V^S_{E,\vec{n},\lambda}(t,\vec{x})\nonumber\\
&&~~~=iN e^{iEt}\int_{0}^{\infty} s\, ds \left(
\begin{array}{c}
-\lambda\,e^{-\pi k/2}H^{(2)}_{\nu_{-}}(s)\,
\eta_{\lambda}(\vec{n})\\
\frac{1}{2}\,e^{\pi k/2}H^{(2)}_{\nu_{+}}(s) \,\eta_{\lambda}(\vec{n})
\end{array}\right)
e^{-i \omega s \vec{n}\cdot\vec{x}+i\epsilon \ln s}\,,\label{Vps}
\end{eqnarray}
where we denote the dimensionless integration variable by $s=\frac{p}{\omega}$
and take
\begin{equation}\label{norm}
N=\frac{1}{(2\pi)^{3/2}}\, \frac{\omega}{\sqrt{2}}\,.
\end{equation}
Then it is not hard to verify that these spinors are charge-conjugated to each
other,
\begin{equation}\label{conj}
V^S_{E,\vec{n},\lambda}=(U^S_{E,\vec{n},\lambda})^{c}={\cal C}
(\overline{U}^S_{E,\vec{n},\lambda})^T \,, \quad {\cal C}=i\gamma^2\gamma^0\,,
\end{equation}
and satisfy the orthonormalization relations
\begin{eqnarray}
&&\left<U^S_{E,\vec{n},\lambda},U^S_{E,\vec{n}^{\,\prime},\lambda^{\prime}}\right>=
\left<V^S_{E,\vec{n},\lambda},V^S_{E,\vec{n}^{\,\prime},\lambda^{\prime}}\right>\nonumber\\
&&\hspace*{40mm}= \delta_{\lambda\lambda^{\prime}}\delta(E-E')\,\delta^2
(\vec{n}-\vec{n}^{\,\prime})\,,
\label{orto1}\\
&&\left<U^S_{E,\vec{n},\lambda},V^S_{E,\vec{n}^{\,\prime},\lambda^{\prime}}\right>=
\left<V^S_{E,\vec{n},\lambda},U^S_{E,\vec{n}^{\,\prime},\lambda^{\prime}}\right>=
0\,.\label{orto2}
\end{eqnarray}
deduced as in the Appendix B. However, the most important result is that in SP
these fundamental spinors  accomplish the condition
\begin{eqnarray}\label{compl}
&&\int_0^{\infty}dE\int_{S^2} d\Omega_n \sum_{\lambda}\left\{
U^S_{E,\vec{n},\lambda}(t,\vec{x})[U^{S}_{E,\vec{n},\lambda}(t,\vec{x}^{\,\prime})]^+
\right.\nonumber\\
&&~~~~~~~~~~~~~~~~~\left.+\,V^S_{E,\vec{n},\lambda}(t,\vec{x})
[V^{S}_{E,\vec{n},\lambda}(t,\vec{x}^{\,\prime})]^+ \right\}=\delta^3
(\vec{x}-\vec{x}^{\,\prime})\,,
\end{eqnarray}
which means that they form a {\em complete} system of solutions.

The last step is to rewrite all these results in  NP where the Dirac field,
\begin{eqnarray}
\psi(x)=\psi_S(t,e^{\omega t}\vec{x})=\int_0^{\infty}\,dE\int_{S^2}\,
d\Omega_n\,\,
\sum_{\lambda}\left[U_{E,\vec{n},\lambda}(t,\vec{x}) a(E,\vec{n},\lambda)\right.~~~&&\\
\left.+\, V_{E,\vec{n},\lambda}(t,\vec{x}) b^*(E,\vec{n},\lambda)\right]\,,&&
\end{eqnarray}
depends on the fundamental solutions in NP,
\begin{equation}
U_{E,\vec{n},\lambda}(t,\vec{x})=U^S_{E,\vec{n},\lambda}(t,e^{\omega
t}\vec{x})\,,\quad
V_{E,\vec{n},\lambda}(t,\vec{x})=V^S_{E,\vec{n},\lambda}(t,e^{\omega
t}\vec{x})\,,
\end{equation}
which satisfy orthonormalization relations similar to Eqs. (\ref{orto1}) and
(\ref{orto2}) but a different completness relation,
\begin{eqnarray}\label{compl1}
&&\int_0^{\infty}dE\int_{S^2} d\Omega_n \sum_{\lambda}\left\{
U_{E,\vec{n},\lambda}(t,\vec{x})[U_{E,\vec{n},\lambda}
(t,\vec{x}^{\,\prime})]^+\right.\nonumber\\
&&~~~~~~~~~~~~~~~~~\left.+\,V_{E,\vec{n},\lambda}(t,\vec{x})
[V_{E,\vec{n},\lambda}(t,\vec{x}^{\,\prime})]^+ \right\}=e^{-3\omega t}
\delta^3 (\vec{x}-\vec{x}^{\,\prime})\,,
\end{eqnarray}
similar to that of Ref. \cite{cot}.

An interesting problem is to find the set of commuting operators which
determine these quantum modes. Apart from the Hamiltonian operator $H$ we must
take into account the conserved operator $\vec{\cal N}$ of the momentum
direction, defined in the momentum representation as $({\cal
N}^i\hat\psi_S)(\vec{p})= n^i \hat\psi_S(\vec{p})$, and the operator ${\cal
W}=\vec{\cal N}\cdot \vec{S}$ which is a  version of the Pauli-Lubanski
operator. The set of fundamental spinors  (\ref{Ups}) and (\ref{Vps}) are
common eigenspinors of the set of commuting operators $\{H,\vec{\cal N}, {\cal
W}\}$. More specific, the concrete eigenvalue problems in NP read
\begin{eqnarray}
&H\,U_{E,\vec{n},\lambda}= E\, U_{E,\vec{n},\lambda}\,, \quad&
H\,V_{E,\vec{n},\lambda}=-E\, V_{E,\vec{n},\lambda} \,,\label{HUV}\\
&{\cal N}^i\,U_{E,\vec{n},\lambda}= n^i\, U_{E,\vec{n},\lambda}\,, \quad&
{\cal N}^i\,V_{E,\vec{n},\lambda}=-n^i\, V_{E,\vec{n},\lambda} \,,\label{PUV}\\
&{\cal W}\,U_{E,\vec{n},\lambda}= \lambda U_{E,\vec{n},\lambda}\,, ~~\quad&
{\cal W}\,V_{E,\vec{n},\lambda}=-\lambda V_{E,\vec{n},\lambda}\,.\label{WUV}
\end{eqnarray}

Hance we get all the ingredients we need for introducing the second quantization
in canonical manner as in Ref. \cite{cot} and calculate the conserved operators
of the quantum field theory. Let us observe that the wave functions $a$ and $b^*$ (which
have to become field operators) are the same in both the picture considered here which means
that the second quantization is independent on the picture choice. A further paper
will be devoted to this problem.

\section{Concluding remarks}

We derived new Dirac quantum modes in moving frames of the dS spacetime using
the SP which allows us to separate the coordinate or momentum variables of the
free Dirac equation. This new picture is related to the NP through a
non-unitary transformation which preserves the eigenvalues equations but
changes the form of the relativistic scalar product of the SP, eliminating from
Eq. $(\ref{SP2})$ the weight function $\sqrt{g(t)}$ of the original scalar
product $(\ref{SP1})$. However, this is not an impediment since the
relativistic scalar product has no more the same immediate physical
interpretation as that of the non-relativistic quantum mechanics. Moreover, we
believe that the elimination of this weight function is a remarkable advantage
since in this way the scalar product $(\ref{SP2})$ becomes just that of special
relativity.  Nevertheless, if one does not agree with these arguments then one
can use our SP only as a tool for finding new solutions, following to carry out
the results in NP where the physical interpretation is evident. Anyway, it is
obvious that this new picture helps one to derive new quantum modes which never
could be found in NP.

The principal result obtained here is the complete system of orthonormalized
energy eigenspinors determined in dS moving charts by the set of commuting
operators $\{H,\vec{\cal N}, {\cal W}\}$. We remind the reader that in Ref.
\cite{cot} we found another set of Dirac solutions  which form a complete
system of orthonormalized  eigenspinors of the commuting operators $\{\hat
P^i,{\cal W}\}$ in the same dS chart. Since $H$  does not commute with $\hat
P^i$ it results that these two sets of quantum modes are quite different.
Therefore, we have now a complete theory of free Dirac quantum modes in dS
moving frames helping us to understand how can be measured the momentum in
states of given energy and, reversely, the energy in states having a
well-determined momentum.  We hope that in this way a coherent relativistic
quantum mechanics has to be built as the starting point to a successful theory
of Dirac quantum fields on dS backgrounds.

Finally, we note that the quantum theory on dS manifolds we try to develop
takes into account quantum modes {\em globally} defined as eigenstates of
different sets of commuting conserved operators. These can be chosen from the
large algebra of conserved observables produced by the high symmetry of the dS
geometry. Physically speaking this means that we use a global apparatus
providing the same type of measurements on the whole domain of the observer's
chart. In our opinion, this attitude does not contradict the general concept of
local measurements \cite{BD} which is the only possible option when the
symmetries are absent and the global apparatus does not work.

\subsection*{Acknowledgments}

We are grateful to Mihai Visinescu for interesting and useful discussions  on
closely related subjects.

\appendix

\subsection*{Appendix A: Some properties of Hankel functions}

According to the general properties of the Hankel functions
\cite{AS}, we deduce that those used here,
$H^{(1,2)}_{\nu_{\pm}}(z)$, with $\nu_{\pm}=\frac{1}{2}\pm i k$
and $z\in \Bbb R$, are related among themselves through
\begin{equation}\label{H1}
[H^{(1,2)}_{\nu_{\pm}}(z)]^{*} =H^{(2,1)}_{\nu_{\mp}}(z)\,,
\end{equation}
satisfy the equations
\begin{equation}\label{H2}
\left(\frac{d}{dz}+\frac{\nu_{\pm}}{z}\right)H^{(1)}_{\nu_{\pm}}(z)
=  i e^{\pm \pi k} H^{(1)}_{\nu_{\mp}}(z)
\end{equation}
and the identities
\begin{equation}\label{H3}
e^{\pm \pi k} H^{(1)}_{\nu_{\mp}}(z)H^{(2)}_{\nu_{\pm}}(z) +
e^{\mp \pi k}
H^{(1)}_{\nu_{\pm}}(z)H^{(2)}_{\nu_{\mp}}(z)=\frac{4}{\pi z}\,.
\end{equation}

\subsection*{Appendix B: Normalization integrals}

In spherical coordinates of the momentum space, $\vec{n}\sim
(\theta_n,\phi_n)$, and the notation $\vec{p}=\omega s\vec{n}$, we have $d^3p=
p^2dp\, d\Omega_n=\omega^3\, s^2ds\, d\Omega_n$ with
$d\Omega_n=d(\cos\theta_n)d\phi_n$. Moreover, we can write
\begin{equation}\label{del}
\delta^3(\vec{p}-\vec{p}^{\,\prime})=\frac{1}{p^2}\,\delta(p-p')\delta^2(\vec{n}-\vec{n}')
=\frac{1}{\omega^3 s^2}\,\delta(s-s')\delta^2(\vec{n}-\vec{n}')\,,
\end{equation}
where we denoted $\delta^2(\vec{n}-\vec{n}') =\delta(\cos \theta_n-\cos
\theta'_n)\delta(\phi_n-\phi'_n)\,.$

Then the scalar products of the fundamental spinors of positive frequencies can
be calculated according to Eqs. (\ref{Ups}), (\ref{norm}), (\ref{H3}),
(\ref{Porto}) and (\ref{del}) as
\begin{eqnarray}
\left<U^S_{E,\vec{n},\lambda},U^S_{E,\vec{n}^{\,\prime},\lambda^{\prime}}\right>=
\int_D d^3x\, [{U}^S_{E,\vec{n},\lambda}(t,\vec{x})]^+
U^S_{E,\vec{n}^{\,\prime},\lambda^{\prime}}(t,\vec{x})
&&\nonumber\\
= e^{i(E-E')t}\left[\frac{1}{2\pi\omega}\int_0^{\infty}\frac{ds}{s}\,
e^{i(\epsilon-\epsilon')\ln s}\right] \delta_{\lambda\lambda^{\prime}}\,
\delta^2
(\vec{n}-\vec{n}^{\,\prime})&&\nonumber\\
=\delta_{\lambda\lambda^{\prime}}\delta(E-E')\,
\delta^2(\vec{n}-\vec{n}^{\,\prime})\,.~~~~~~~~~~~~~~~&&
\end{eqnarray}
The properties (\ref{orto1}) - (\ref{compl}) are deduced in the same manner.

\end{document}